\documentclass[a4paper]{article}

\usepackage[pages=all, color=black, position={current page.south}, placement=bottom, scale=1, opacity=1, vshift=5mm]{background}

\usepackage[margin=1in]{geometry} 

\usepackage[T1]{fontenc}

\usepackage{amsmath}
\usepackage{amsthm}
\usepackage{amssymb}

\usepackage[utf8]{inputenc}
\usepackage{hyperref}
\hypersetup{
	unicode,
	pdfauthor={Author One, Author Two, Author Three},
	pdftitle={A simple article template},
	pdfsubject={A simple article template},
	pdfkeywords={article, template, simple},
	pdfproducer={LaTeX},
	pdfcreator={pdflatex}
}

\usepackage[sort&compress,numbers,square]{natbib}
\theoremstyle{plain}

\theoremstyle{definition}

\usepackage{graphicx, color}
\graphicspath{{fig/}}

\usepackage{algorithm, algpseudocode} 
\usepackage{mathrsfs} 

\usepackage{lipsum}

\title{Ensemble CNN and Uncertainty Modeling to Improve Automatic Identification/Segmentation of Multiple Sclerosis Lesions in Magnetic Resonance Imaging}
\author{Giuseppe Placidi$^1$ \and Luigi Cinque$^2$ \and Daniela Iacoviello$^3$  \and Filippo Mignosi$^4$ \and Matteo Polsinelli$^1$}

\date{
	$^1$ A2VI-lab, c/o Department of MeSVA, University of LAquila, Via Vetoio Coppito 2, 67100 - LAquila, ITALY 
	\\ \texttt{giuseppe.placidi@univaq.it}\\%
	
	$^2$ Department of Computer Science, Sapienza University of Rome, Via Salaria 113, 00198 - Roma, ITALY \\ 
	
	$^3$ Department of Computer, Control and Management Engineering Sapienza University of Rome, Rome, Italy
	
	$^4$ Department of DISIM, University of LAquila, Via Vetoio Coppito, 67100 - LAquila, ITALY\\
}

\begin{document}
	\maketitle
	
\begin{abstract}
To date, several automated strategies for identification/segmentation of Multiple Sclerosis (MS) lesions with the use of Magnetic Resonance Imaging (MRI) have been presented, but they are outperformed by human experts, from whom they act very differently. This is mainly due to: the ambiguity originated by MRI instabilities; peculiar variability of MS; non specificity of MRI regarding MS. Physicians partially manage the uncertainty generated by ambiguity relying on radiological/clinical/anatomical background and experience. 
To emulate human diagnosis, we present an automated framework for identification/segmentation of MS lesions from MRI based on three pivotal concepts: 1. the modelling of uncertainty; 2. the proposal of two, separately trained, CNN, one optimized for lesions and the other for lesions with respect to the environment surrounding them, respectively repeated for axial, coronal and sagittal directions; 3. the definition of an ensemble classifier to merge the information collected by different CNN.

The proposed framework is trained, validated and tested on the 2016 MSSEG benchmark public data set from a single imaging modality, the FLuid-Attenuated Inversion Recovery (FLAIR). The comparison with the ground-truth and with each of 7 human raters, proves that there is no significant difference between the automated and the human raters.

\end{abstract}

\section{Introduction}
\label{sec1}

Multiple Sclerosis (MS) is a degenerative disease mainly affecting white matter (WM) and spinal cord, with a very heterogeneous clinical presentation across patients both in severity and symptoms \cite{Stein1996}. 

The origins of the disease are not well understood but characteristic signs of tissue degeneration are recognizable by the presence of lesions and brain atrophy. Most of MS signs can be observed through Magnetic Resonance Imaging (MRI) which has become the elective mini-invasive tool for MS  \cite{Filippi2019}. Focal lesions 
are primarily visible in the white matter on structural MRI, observable as hyper-intensities on T$_2$-weighted images (T$_2$w), on proton-density images (PD), or on FLuid-Attenuated Inversion Recovery images (FLAIR), and as hypo-intensities on T$_1$-weighted images (T$_1$w). 
%
%

Physicians often use FLAIR images for WM lesion detection and other modalities mostly to ascertain the presence of cortical lesions. 

An examination consists in thousands of images mostly collected pre and post contrast administration. MRI is used routinely in clinical practice but it is unspecific for MS and not well correlated to the impairment progression, to the neuro-plasticity and to the effects of demyelinization of nerves, the last being a critical effect which is invisible to MRI. In a MS patient, the "healthy" brain tissue is usually referred as "apparently healthy" \cite{Filippi2019}. In addition, often lesions and healthy tissue are present in the same place, thus resulting in the partial volume effect (PVE). Regarding MRI, healthy anatomical structures similar to lesions and close to lesions could contribute to create further ambiguity. 
Moreover, there is a huge variability in images due to differences in scanners, magnetic field strength/homogeneity and tuning of parameters  \cite{Placidi2012}. Efforts have been made to standardize MRI amplitude \cite{Carr2020} but results are still unsatisfactory due to the huge set of parameters to be tuned. 
The consequence of the above phenomenons are ambiguities and disagreement among radiologists (inter-raters variability) as well as uncertainty in a same radiologist (intra-rater variability) mainly in defining the borders of the lesions, but also in labeling whole regions. This could make the manual segmentation, besides long and boring, also inaccurate, especially when quantitative evaluations are required, despite the ability of radiologists in relying on unexpressed anatomical/clinical background, on knowledge about MS and its symptoms.

Several automatic frameworks have been recently proposed and reviewed \cite{Danilakis2018, Commowick2018, 
Placidi2019,Zhang2020, ghosal2019light,alijamaat2021multiple,gros2021softseg}. 
Despite the huge efforts, to date the results are still far from those of human experts. Actually, this has led to an increase in the model complexity not corresponding to the expected improvement \cite{mckinley2021simultaneous}. 
Automated strategies are not robust to MRI variability, not even sufficiently able to model medical knowledge, human operational capacity and flexibility. 
Furthermore, the reasoning methodology used by radiologists for volume analysis, consisting in 2D axial slices with a continuous view of coronal and sagittal slices for confirmation \cite{traboulsee2016revised,Filippi2019},
is not fully imitated by automatic strategies.
A recent paper \cite{Macar2021} highlights the importance of using 3D CNN in MS lesion segmentation, though other recent frameworks \cite{Vaanathi2021,Yang2021} prefer ensembles of 2D U-net models, in a way which is similar to the human methodology. Human methodology is empowered by the fact that spatial resolution is often not the same along the three axes with a prevalence for axial planes. 
However, the uncertainty affecting radiologists during classification is not reported in the public data sets: a binary choice is often insufficient to represent the evaluation of an expert. If represented, the uncertainty could greatly help an automatic strategy to better segment also undoubted lesions. This pushed to investigate on uncertainty in medical data \cite{Alizadehsani2021} 
and on the effect that the rater style could transfer in terms of uncertainty to an automated strategy \cite{Vincent2021}.

Implicit information in automated methods is difficult to be modelled and more, if introduced with external supports (dictionary, anatomical atlases, etc.) \cite{Chengao2021}, is often insufficient to fill the gap with human experts. 

We aim at filling this gap, both in performance and in reasoning, by proposing a framework which includes: 
1) the classification of uncertainty as an intermediate class between the background and lesions; 2) the optimization of two CNN (2D U-net models), one for the class lesion and one for the class background to contextualize lesions with respect to the surrounding anatomical structures for the three spatial directions (axial, coronal and sagittal); 
3) the definition of an ensemble classifier to merge the information collected by all CNN. 

To obtain our goal we: use the publicly available large-scale benchmark MRI database and corresponding ground truth proposed in 2016 MICCAI MS Lesion Segmentation Challenges, MSSEG \cite{Commow2016}; define the uncertain regions with the help of the binary classifications of 7 human raters in MSSEG; apply the framework just on FLAIR images. 

These choices allow to: compare the proposed framework with competitive automated strategies and human experts; demonstrate that the uncertain reasoning modelling could help in reducing the ambiguity and complexity of the, intrinsically uncertain, problem; demonstrate that a single imaging sequence, FLAIR, is sufficient for MS lesion identification/segmentation in WM. 
It is implicit that the framework applies to WM lesions and not on cortical lesions, being these last absent in MSSEG.
 


The manuscript is structured as follows: Section 2 presents a review of automatic approaches to MS lesion identification/segmentation, in particular those  using CNN; Section 3 describes the proposed framework in the context of the used data set, the defined three-class consensus used for training, the proposed CNN architecture and the ensemble system; Section 4 details the metrics used for comparison; Section 5 reports and discusses experimental results; Section 6 concludes the paper and presents some constructive hints for future improvements. 



\section{Related work}
Medical image analysis is greatly performed with automated methods, mostly involving deep learning  \cite{Litjens2017,Dong2019,Mingxia2020,Jie2021,Chunfeng2022}. Automated MS lesion identification$/$segmentation is still an active field of research and several methods have been provided in the last decade and well reviewed along time 
\cite{Danilakis2018,Commowick2018,Kaur2020,Zhang2020} 
with an emerging role of AI-based methods \cite{Afzal2020}.


Automated strategies can be classified into three main groups: methods using pre-selected features modelling (PSFM), methods using a-priori information modelling (APIM) and methods using deep learning modelling (DLM).

PSFM calculate pre-selected features and learn from previously segmented training images to separate lesions from healthy tissue \cite{Zurita2018}. Some PSFM use a large set of features and select the more discriminant ones through labelled training. Others use an atlas-based technique, employing topological and statistical atlases for WM lesion segmentation \cite{Shiee2010,Huisi2021}; another includes the usage of Decision Random Forests \cite{Geremia2011}.
Similarly,  a framework for segmentation of contrast-agent enhanced lesions using conditional random fields is defined in \cite{Karimag2012}. 
The work in \cite{Cabezas2014} proposes a set of features, including contextual features, registered atlas probability maps and an outlier map, to automatically segment MS lesions through a voxel by voxel approach. A rotation-invariant multi-contrast non-local means segmentation is proposed in \cite{Guizard2015}  for the identification and segmentation of lesions from 3D MRI images. Supervised learning by PSFM has been widely employed in tasks where the training database and the pre-selected feature set cover all possible cases \cite{Carass2017}. Nevertheless, when the heterogeneity of the disease and the potential variability of imaging are large, as it occurs for MS and MRI, the dimension of the training database and, mostly, the choice of the pre-selected features are critical.

APIM does not require labelled training data to perform segmentation, 
but usually exploit some a-priori information, such as the intensity clustering\cite{Yoo2014}, to model tissue distribution. 
In \cite{Ali2005}, a likelihood estimator to model the distribution of intensities in healthy brain MR images is presented. Other methods use threshold with post processing refinement 
\cite{Roura2015}
or are based on probabilistic models 
\cite{Strumia2016}. A big challenge for APIM is that the outliers are not specific for lesions because they could be due to artifacts, intensity inhomogeneity and small anatomical structures like blood vessels: this often produces false positives \cite{Brosh2016}. Moreover, APIM is strongly based on the information extracted and simplified by the knowledge of specific experts.

During the last years DLM has gained popularity in medical imaging especially with 
U-nets ad their variants \cite{Ronneb2015,Iek2016,Milletari2016,blabla2021}. Though the dimension of the training database is also crucial in DLM, this has no concern regarding the pre-selection of features as in PSFM or regarding a-priori information modelling as in APIM.
In particular CNN, compared to machine learning approaches, has achieved remarkable success in biomedical image analysis
\cite{Brosh2016,Dong2019,Jie2021,Yang2021}. DLM trains and learns to design features directly from data \cite{Aslani2019} and provides best results in MS lesion identification/segmentation, as recently confirmed in \cite{Danilakis2018,Commowick2018,Kaur2020,Zhang2020,kang2020acu, vang2020synergynet, zhang2021geometric, gros2021softseg, alijamaat2021multiple}. 


CNN applied to MS often use 2D spatial convolutional layers \cite{Aslani2019}, others use 3D convolutional layers to incorporate  3D spatial information simultaneously \cite{Valverde2017,Hashemi2019,Larosa2019,Valverde2019} or merge spatial with temporal information 
\cite{Placidi2019}. All these methods perform segmentation with a minimum lesion volume threshold to avoid the inclusion of small false positives.


Some methods improve the MS segmentation by retraining the CNN from the first layers, in a kind of self supervision \cite{Fennetau2020}, by hypothesizing that the those layers are more sensitive.

Among other recent works, hybrid models \cite{Elsebely2021} allow to detect MS lesions by using textural features in combination with machine learning approaches, as Decision Tree (DT) and Support Vector Machines (SVM). 

However, CNN performance is still far from that obtained by human experts or dramatically drops with other data sets \cite{mckinley2021simultaneous}.


\section{The proposed framework}

\begin{figure*}[!t]
\centering
\includegraphics[scale=0.5]{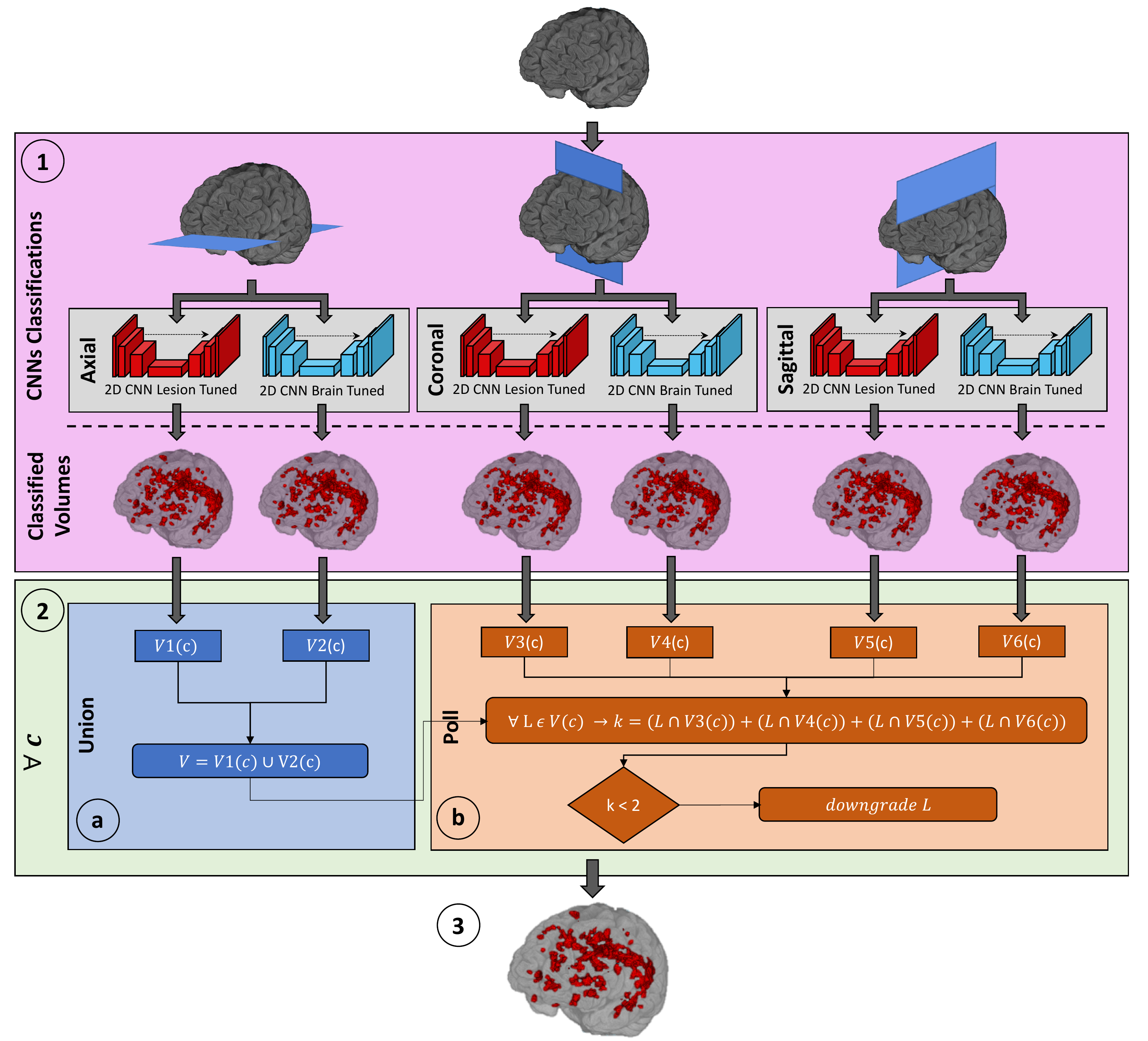}
\caption{\label{fig01}
The proposed identification/segmentation pipeline which divides the brain tissue in three classes: healthy tissue (Background), tissue that has uncertain nature (Uncertainty), and MS lesions (Lesion). The strategy operates independently on axial, coronal and sagittal images, each processed by two separately trained U-nets, one optimized for Lesion, to directly focus on lesions, and the other optimized for contextualizing lesions with respect to the environment. It recombines the results by using the Union of axial volumes followed by a majority vote strategy on the coronal and sagittal volumes, for confirmation. Voxels whose classification is not confirmed are downgraded (Lesion becomes Uncertainty and Uncertainty becomes Background). The framework operates separately for Lesion and Uncertainty, starting from Lesion. In step 2b, the procedure is applied voxel by voxel: $L$ is referred to each single voxel of the class $c\in \{Lesion, Uncertainty\}$.}
\end{figure*}

The framework we propose, sketched in Figure \ref{fig01}, consists of the following steps: 1) deep learning automatic classification, of the images (2D) composing the MRI model, in three classes: Background, Uncertainty and Lesion (capital letter to imply the concept 'class'),  optimized for Lesion (lesions from inside) and for Lesion in the context of the surrounding environment, separately for axial, coronal and sagittal directions (resulting in 6 classifiers); 2) class fusion (separately for Lesion and Uncertainty, starting from Lesion) by performing the Union of the 2 axial segmentation (step 2a in Figure \ref{fig01}), followed by a majority vote taken from the remaining segmentations and used for confirmation of the class (if the class is not confirmed, this is downgraded (step 2b in Figure \ref{fig01}); 3) final output. 

For the framework we propose, the following three hypotheses hold: 

1) the MSSEG pre-processed data from just one single MRI modality, FLAIR, are the input of the framework;

2) the binary labelled ground-truth is revised to contain, besides Lesion and Background, also Uncertainty which is created from part of the original Background, leaving Lesion unchanged (see detailed description below);

3) the three considered classes are supposed to be ordered, Background $<$ Uncertainty $<$ Lesion: in our case, just Lesion and Uncertainty are the subjects of fusion and, for this reason, their downgrading consists in the passage from Lesion to Uncertainty and from Uncertainty to Background, respectively: the process starts from Lesion to allow Uncertainty fusion on the upgraded data set.

Step 2a of Figure \ref{fig01} serves to include, besides common information, also complementary ones coming from the specificity of each of the two axial CNN and, at the same time, to model the reasoning of the radiologists that use axial orientation to make the first hypotheses.  Step 2b is used to vote for each object resulting from the axial processing (Lesion or Uncertainty), its permanence in the assigned class, or its downgrading. 
Objects are confirmed when at least two of the other four raters (two coronal and two sagittal) agree with the axial classification. This ensures that false positives are greatly reduced, that 3D contextualization with the environment is maintained and that the model agrees with the radiologist's reasoning. 

The choices regarding the usage of one single imaging modality, the classification in three classes and the use of an ensemble framework are clarified below.

Being supervised, each classifier needs training, validation and test carried on by using data from a public data set. In what follows we first describe the used data set, the ternary ground truth, the CNN architecture, the used loss function, the hyper-parameters optimization and the ensemble, final, classification of Figure \ref{fig01}.


%

\subsection{MSSEG data set}\label{subsecMSSEG}


To allow a direct comparison and benchmark of the proposed framework with the state-of-the-art segmentation methodologies, we use the 2016 MSSEG data set \cite{Commow2016}. In MSSEG, data were collected with several MRI scanners using different magnetic field strengths: 1.5T Siemens Aera, 3T Siemens Verio, 3T Philips Ingenia and 3T General Electric Discovery. The data set contains, for each examination, $T_1$-w, $T_1$-w gadolinium enhanced ($T_1$-w Gd), $T_2$-w, $T_2$-FLAIR and $PD$-w images.

We have chosen MSSEG because it is a benchmark data set including 7 different classifications made by 7 different expert human raters and a consensus, the ground-truth, making it possible to compare our framework with human classifications, as well as with state of the art automatic strategies.
For MSSEG, it has been asked the human raters to perform a binary segmentation (each voxel was identified as lesion/not lesion): the annotated data set has been composed by 15 training cases. Another subset of data, composed of 38 testing cases, has been left without annotation. 


The images are anonymized and provided both in original form and pre-processed for user convenience. 

Pre-processing refers to a series of mathematical adjustments to MR images before segmentation \cite{Piclin2018} for reducing the effects of noise and imaging artifacts, equalizing space, eliminating outliers and stabilizing the contrast. 

Though the MSSEG data set contains images from 4 imaging modalities, in our framework we opt for the usage of one single modality (FLAIR) to: detect/segment MS lesions in WM; demonstrate that a single modality is sufficient for MS lesion identification/segmentation in WM; avoid that the potential gain of using multiple imaging modalities could be overcome by the drop in robustness and performance due to residual MRI instabilities after pre-processing.    


Further, we divide the annotated MSSEG data set, composed by 15 subjects, in three subsets respectively for training, validation and test, as follows:
\begin{itemize}
    \item the training data set contains examinations from 9 subjects, 3 subjects of each centre;
\item the validation data set contains data from 3 subjects, one for each centre;
\item the test data set contains the remaining data, from 3 subjects, one for each centre. 
\end{itemize}

An exhaustive cross validation is performed while maintaining the proportions between centres. Once the data set is established, it is 
augmented by adding, for each image, 2 random rotations (between -13 and 13 degrees, 1 degree in resolution), 1 random scaling (between 1.1 and 1.3, 0.01 in resolution) and 1 gaussian random noise addition, 0 mean and 0.001A variance, where A is the maximum amplitude value in the examined volume.

The augmented data set, for each validation and orientation (axial, coronal and sagittal), contains 5216 images for training, 418 images for validation and 435 images for test.

The augmented data originated from the same dataset \cite{Commow2016} used to train, validate and test the state of the art methods. In this way, we can indirectly compare our framework with the state of the art strategies by using, for them, the performance evaluation reported in the original manuscripts.

\subsection{Ternary ground-truth}

In medical imaging it can be assumed that there is a single, unknown, true segmentation map of the underlying anatomy, and each human rater produces an approximation  with variations reflecting individual experience. 

In the opposite case, it can be assumed that the variable annotations from experts are all realistic and acceptable instances of the true segmentation. 

As it often occurs, the truth is in the middle: some ambiguities are indeed due to human subjectivity or imperfections (extrinsic), while some others are due to the problem itself (intrinsic). In our problem, both are important, but intrinsic ambiguities have the highest role, being due both to MS presentation and to MRI non specificity: lesions are not well separated from healthy tissue in MS (due to PVE) and MRI is neither sufficiently specific for MS nor sufficiently precise. Regarding human subjectivity, it produces differences that are due to a mix of prior assumptions, like experience in the field, greater or lesser exploitation of additional meta-information (such as anatomical/radiological/clinical knowledge), mistakes or oversights which often are concentrated on small and/or low intensity lesions and lesion borders. 

When raters are forced to provide a binary segmentation, as in MSSEG, they cannot express any doubt, whatsoever is the cause. The binary segmentation does not allow the representation of the intrinsic uncertainty and, furthermore, induces a human rater to assume polarized decisions which, from one side, could not correspond to what the rater really believes in and, from the other side, could be confusing and misleading for an automated strategy. In fact, ambiguous decisions might have been assumed by the rater in similar situations (an uncertain region could be considered healthy tissue in one case and lesion in another) which could influence the automated strategy \cite{Vincent2021}. 

If we want to train an automated method to recognize the problem-specific uncertainty, we have to integrate the binary ground-truth with human uncertainty (doubts). 
To maintain the original ground-truth, 
we just consider as Uncertainty the voxels that at least three of the seven human raters considered as lesions and the binary consensus had not. In this way, since we create space for the Uncertainty from the Background, the Lesion class of the binary consensus remains unchanged:
this allows the comparison with state of the art strategies, performed on Lesion. The use of Uncertainty in this specific context 
is intended to reduce the uncertainty in identifying Lesion, hence to improve classification. A more accurate definition of the uncertainty, though interesting, is out the scope of the manuscript and deserves specific exploration.

The method we propose is quite different from other strategies used to define the uncertainty \cite{Kats2019,gros2021softseg}, for the following motivations: 
1) to maintain the original structure of Lesion; 
2) to account for the uncertainty affecting both the problem and the raters;
3) to avoid the new class Uncertainty could 
prevent a direct comparison with other methods;
4) to quantify the performance gain the proposed framework could effectively get when the Uncertainty is introduced; 
5) to allow the learning strategy to consider as uncertain not only lesion borders, as other Authors do \cite{Kats2019}, but also whole regions not necessarily connected to lesions. In fact, Uncertainty could regard both the lesion borders, where damaged tissues could coexist with healthy tissues (PVE), and whole structures, where doubts are due to MRI unspecific nature for MS.

\begin{figure}[!t]
\centering
\includegraphics[scale=.31]{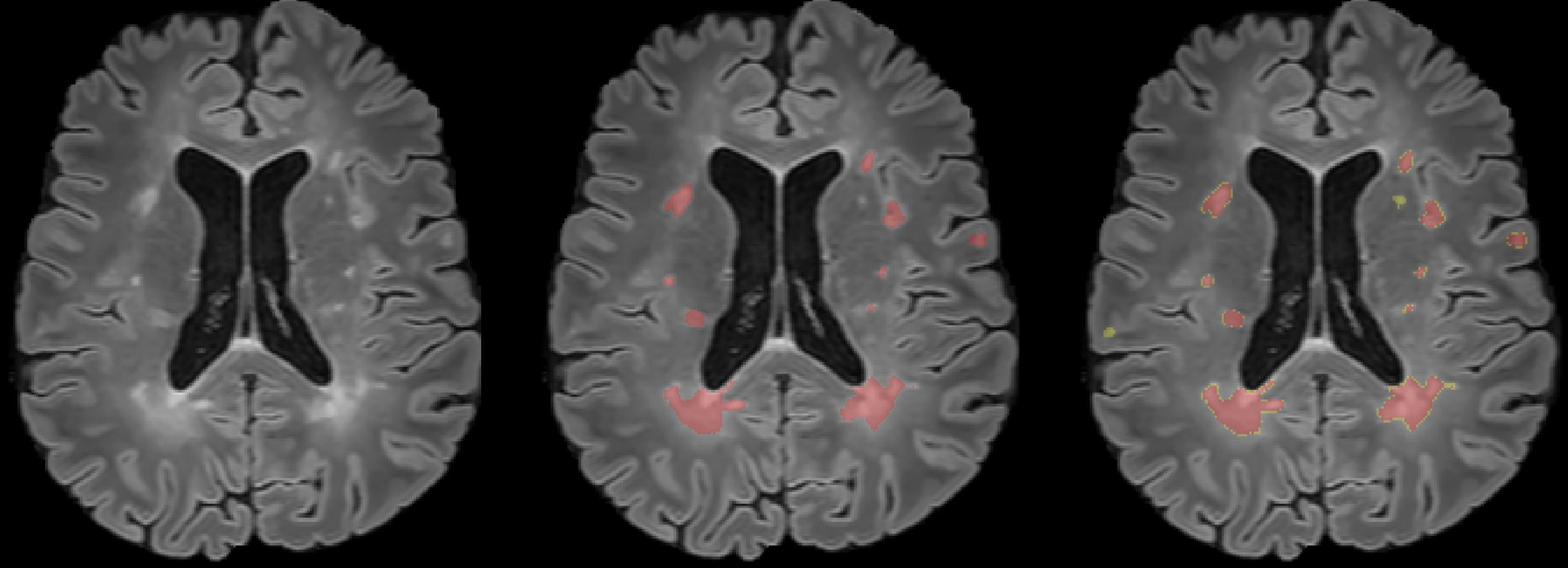}
\caption{\label{fig06}A sample FLAIR image from the MSSEG data set (left), the binary consensus (middle) and the proposed ternary consensus (right). 
The Lesion is annotated in red. In the ternary consensus, the Uncertainty is indicated in yellow.
}
\end{figure}

Fig.\ref{fig06} reports a FLAIR image example with the 
corresponding binary consensus and the proposed ternary consensus. 
Uncertainty is in yellow in the ternary consensus, 
the last used as ground truth to train our framework in reaching a ternary classification.

\subsection{CNN architecture}

\begin{figure*}[!t]
\centering
\includegraphics[scale=0.4]{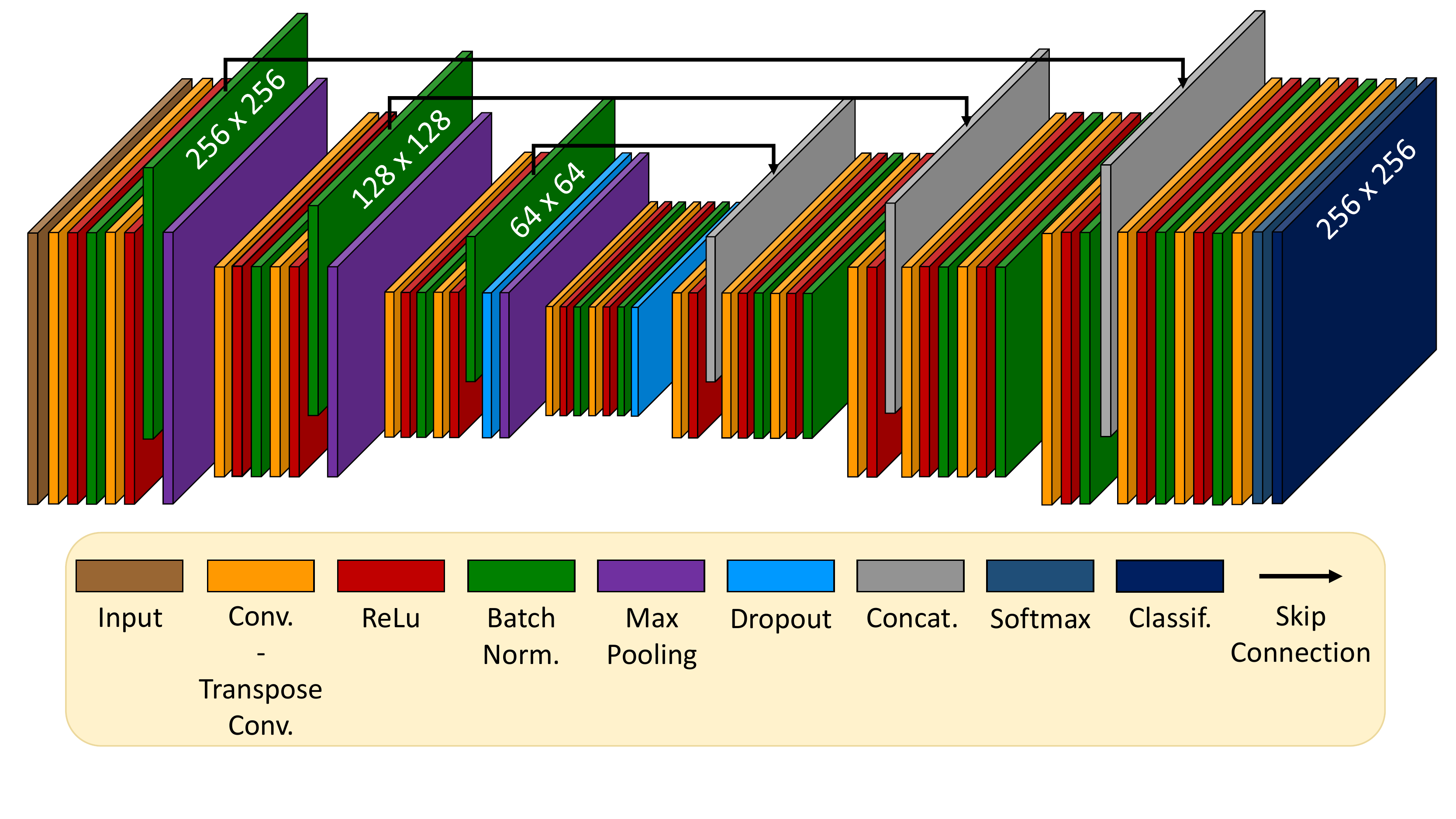}
\caption{\label{figunet} The used U-net "D architecture. The architecture is the same for 6 classifiers, though they have been trained separately.}
\end{figure*}

Being the problem at hand the classification of 2D slices of a FLAIR brain volume into the three classes, Background, Uncertainty and Lesion, in this work we use the U-Net 2D architecture \cite{Ronneb2015} depicted in Figure \ref{figunet}, being U-nets specifically designed for these tasks.

Compared to the traditional U-net architecture, we insert a batch normalization layer in each block to mitigate the effects of the
gradient amplification \cite{Santuk2018} in the regions surrounding the lesions, though with a relevant increase of computational costs (about 30\%). 
To optimize the number of blocks, $n$, we have performed preliminary training, with $n \in \{3, 4, 5 \}$. We have not gone outside this set because for $n = 5$ the U-Net started to overfit, even when using high values of L$_2$-Regularization, and for $n=2$ a dramatic drop of performance occurred. 
With $n = 4$, the problems related to overfitting have disappeared and the performance was good but some redundancy remained. 
For this reason, we have trained the CNN with $n=3$ and verified that redundancy was greatly reduced and training converged faster than for $n=4$: hence, $n=3$ is the number of selected blocks and used thereinafter. 

\subsection{Loss function and process optimization}

The architecture we use has to solve a three class automatic annotation, for which a Multi-label Cross Entropy Loss Function is necessary, defined as follows:

\begin{equation}\label{lossFunction}
loss = \frac{1}{N} \sum_{n=1}^{N} \sum_{i=1}^{K} (T_{ni}log(Y_{ni}) + (1-T_{ni})log(1-Y_{ni}))
\end{equation}

where $N$ and $K$ are the numbers of observations and classes, respectively.

The use of three classes, besides the problem stabilization (the presence of Uncertainty gives better confidence in defining both Lesion and Background), allows also to consider another important aspect. Indeed, we can optimize two CNN, sharing the same architecture and the same loss function (Eq.\ref{lossFunction}), but with a different learning 
focuses: Lesion and Background.  The Uncertainty is used as a sort of "\emph{buffer class}". In the case of a binary classification this would not have been possible: the optimization of one would automatically lead to the optimization of the other (what is not Lesion is Background and vice-versa). The usage of the Uncertainty gave to both CNN a new choice to break that constraint.

To achieve faster training and better performance, in this work the training process of each CNN is controlled by the following hyperparameters 
\cite{snoek2012,Zhang2019p}:

\begin{enumerate}

\item \textbf{Starting Learning Rate}: it is related to the data set and to the type of neural network.
\item \textbf{L2-Regularization}: it prevents overfitting.
\item \textbf{Class balancing}: it optimizes the amplification factor for the represented classes and improves training. Here we have three classes, hence two weights are sufficient (the Lesion weight and the Background one).
\end{enumerate}

Of the above, the first two are standard for CNN, while Class balancing is specific for our CNN because it helps to differentiate the path of optimization. 

The selected hyperparameters are automatically optimized through a Bayesian approach \cite{snoek2012} applied to the following optimization problem:

\begin{equation}
    x^{*} = arg min_{x \in X}  f(x)
\end{equation}

where $X$ is the domain of $x$, $f(x)$ represents an objective function to be minimized and $x^{*}$ is the hyperparameter setting that yields the optimal value of $f(x)$.

In this work $f(x)$ is defined as

\begin{equation}\label{Eqcost}
f(x) = 1- IoU(x)
\end{equation}

where $IoU$ is the Intersection over Union score \cite{Danilakis2018} defined in Section \ref{sec:perf}.

For the two CNN used therein, the $IoU$ is calculated on Lesion for the first and on Background for the second.

\begin{figure}[!t]
\centering
\includegraphics[scale=0.35]{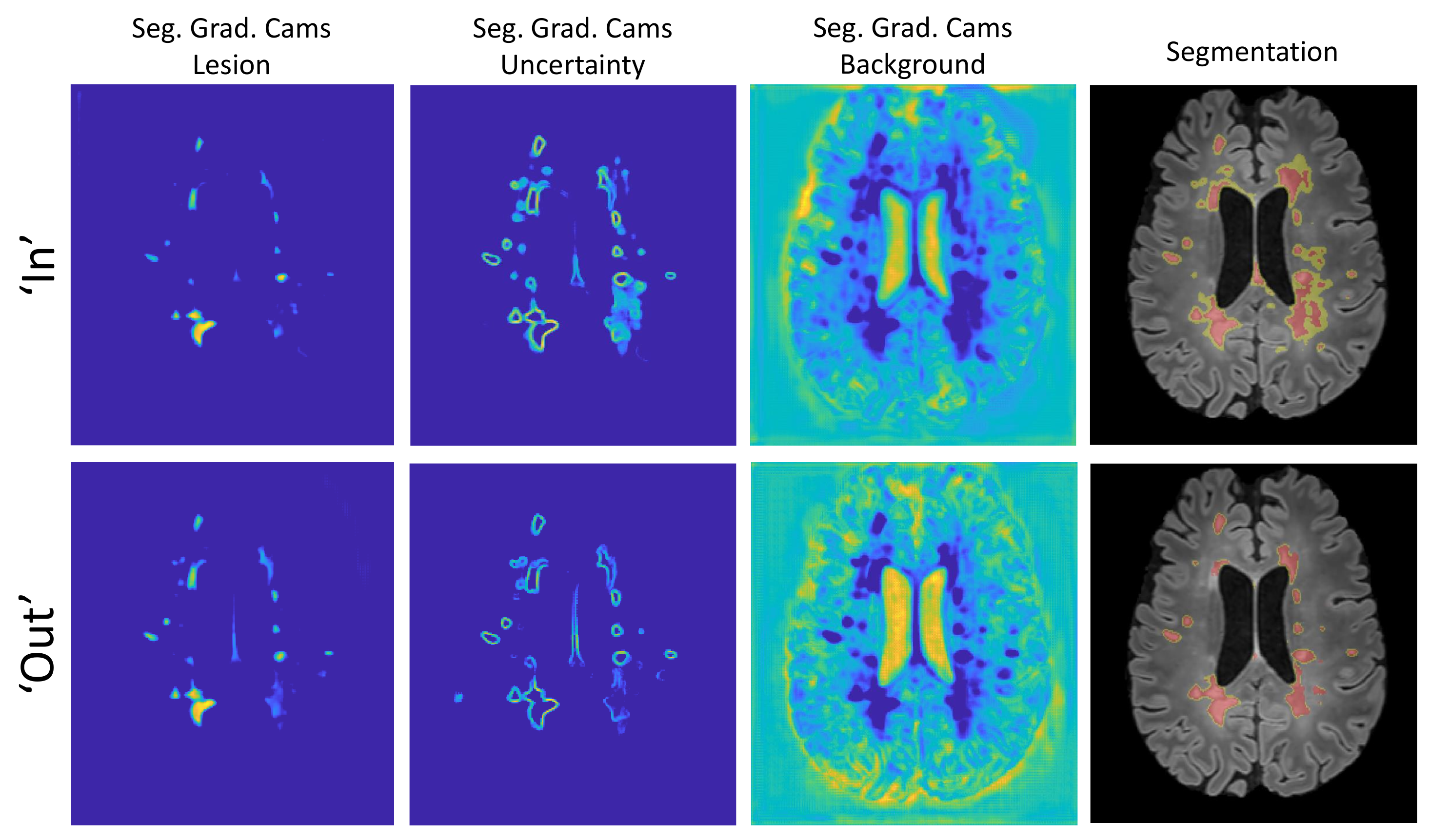}
\caption{\label{figGrad-cam} Grad-cam representation for an axial sample image for both CNN, Lesions from inside (In) and from outside (Out), respectively. Grad-cams are shown for the three classes, Lesion (first column), Uncertainty (second column) and Background (third column). The fourth column shows the resulting classification.}
\end{figure}

The hyperparameter settings for the optimized CNN in each direction are different and justify different training paths for the CNN and different points of convergence for each of them. 
Figure \ref{figGrad-cam} shows the different behaviour of the two CNN in the segmentation grad-cams of a sample image, for the axial direction. The CNN optimized for Lesion, tends to enlarge Lesion and Uncertainty with respect to the CNN optimized for Background, which surrounds lesions from outside.

\subsection{Ensemble Classification}

It is well known that ensemble classifiers often perform better and more robustly than their single components \cite{Yun2021,Vaanathi2021}.
For classification, we use 2D slices of the whole volume, with specific CNN trained separately for Lesion and Background, each for axial, radial and sagittal orientations, respectively. In this way, we avoid that a particular orientation could be favourable to lesions or to the classifier. Further, it serves to ensure 3D continuity and to consider the context of lesions in the surrounding environment.

We obtain a set of 6 classifiers, 
whose classification has to be merged to produce a single output resembling the reasoning of the radiologists: though 3D FLAIR data are collected following sagittal planes to account for clinical/physical issues, radiologists often use axial images for data interpretation and the other orientations for confirmation \cite{traboulsee2016revised,Filippi2019}. Accordingly, we prefer axial classifications and use coronal and sagittal output for confirmation. 

Regarding axial classifications,
since each of the two CNN referring to the same direction operates in the same scenario but with different focuses, they collect specific information where reasoning is specific. Since both specific contributions are important, besides common findings, a Union operation between the two classifications is required. This is in accordance with the double reading procedure followed by radiologists \cite{Geijer2018}. 

However, being the classified volume a three values data set, the Union does not correspond to the classic binary union operation. In our case, the Lesion is privileged, then comes the Uncertainty and, finally, the Background. In fact, a voxel is classified as part of a lesion if at least one of the two classifications considers it as a lesion; elsewhere, if at least one of the two classifications considers it as uncertain, it is classified as Uncertainty, elsewhere, it is considered as Background.

After Union operation, false positives are more present than in each single classifier: their number is reduced by using the majority vote between the other 4 classifications (two coronal and 2 sagittal, being the comparison performed along axial planes). In fact, for each voxel the class is maintained if at least two of the other classifiers confirm it, elsewhere it is downgraded by one (a potential Lesion becomes Uncertainty, a potential Uncertainty becomes Background): a double step is not allowed. 
This means that first a decision is made on the Lesion and, then, on the Uncertainty by using the data set resulting from the application of the process to the Lesion.

The ensemble of different classifiers is justified both by the fact that the Union has to join common information, as well as specific information from each of the two axial classifiers, and because each potentially positive voxel needs confirmation from the coronal and sagittal classifiers (in this case, also 3D spatial information is considered). 
We have preferred to resemble the usual procedure used by radiologists and to rely on the usual benefits of using an ensemble classifier.

In the proposed automatic pipeline, we have copied the human behavior by privileging axial sections with respect to the others, but we have also performed trials regarding the preference of the other orientations in the fusion process and the results, not shown, confirm that the axial preference gives the best results, closely followed by the coronal and, at a great distance, by the sagittal, though this is the direction used for 3D FLAIR data collection. This could be explained, at least partially, by the fact that axial and coronal slices show highly symmetrical shape both regarding brain anatomy and lesion shapes, also across subjects, thus making the learning process easier than for sagittal slices. In sagittal directions symmetry is absent and a huge variation of the image content could correspond to a little rotation of the head.  


\section{Performance Parameters}\label{sec:perf}

In the binary classification problem the voxels can be positive (P, or Lesion) or negative (N or Background). In a ternary ground-truth, P represents the voxels of the class we are considering at present (Lesion or Uncertainty), while N represents the negative voxels (those of the other two classes). According to this definition, the same rules apply for each rater (with respect to the class considered at present): $TP$ are the true positive voxels,
$TN$ are the true negative voxels, $FP$ are the false positive voxels and $FN$ are the false negative voxels. Referring to a given class, a classified voxel can be just in one of the states $TP$, $TN$, $FP$ and $FN$.

As far as we need an exhaustive comparison between all the raters involved therein (artificial, single humans and ground truth), and being a unique performance parameter unavailable, we define and calculate all the mostly known scores and metrics. Despite of a certain redundancy, we use the rule of thumb that it is better to over-describe the results than under-describe them in order to highlight any potential sorts of errors the model could make and to avoid that masking these errors could result in an advantage for the model itself. In what follows, we define all the used parameters by separating scores, defined in $[0,1]$ and whose ideal value is 1, from metrics, defined in $[0,\infty)$ and whose best value is 0. The two groups are distinguished for graphical purposes. For more details about the reported metrics, please refer to \cite{csurka2013good,Commowick2018, Danilakis2018}.

\subsection{Scores (convergent to 1)}

Sensitivity (also called recall or true positive rate) is defined as:

\begin{equation}
SENS = \frac{TP}{TP+FN}
\end{equation}

$SENS$ measures the portion of positive voxels that are correctly identified.
We also distinguish an object sensitivity, $OSENS$, defined as:

\begin{equation}
OSENS = \frac{TP_{o}}{TP_{o}+FN_{o}}
\end{equation}
in which the prefix $O$ and the subscript $o$ indicate we are referring to whole objects and not to single voxels. An object is considered as $TP$ if the intersection with the corresponding object in the ground-truth is not empty.

Specificity ($SPEC$) is defined as: 

\begin{equation}
SPEC = \frac{TN}{TN+FP}
\end{equation}

$SPEC$ represents the portion of negative voxels $N$ that have been correctly identified. For the treated case, where classes are strongly unbalanced, $SPEC$ is biased by the fact that most of the image surface is covered by background (unbalancing).

For the accuracy ($ACC$),
%
%
we use the following normalized definition to reduce unbalancing effects: 

\begin{equation}
ACC = \left(\frac{TP}{TP+FN}+\frac{TN}{TN+FP} \right)/2
\end{equation}


Positive Predicted Value ($PPV$), also called Precision, is defined as:

\begin{equation}
PPV = \frac{TP}{TP+FP}
\end{equation}

$PPV$ represents the portion of voxels identified as positives which are really positives ($TP$). 

As for $OSENS$, we define an object-based $PPV$, $OPPV$ as follows:
\begin{equation}
OPPV = \frac{TP_{o}}{TP_{o}+FP_{o}}
\end{equation}

in which the prefix $O$ and the subscript $o$ indicate we are referring to whole objects not to single objects, as above.


Dice score, also called Sorensen–Dice coefficient, is defined as:

\begin{equation}
	Dice  = \frac{2*TP}{2*TP+FP+FN}
\end{equation}

$Dice$ measures the similarity between two data sets. This index is widely used in AI for the validation of image segmentation algorithms.

$Image$ $Dice$ score uses the same equation of $Dice$ but, while $Dice$ is calculated on the whole data set, $Image$ $Dice$ is applied on each image separately and the results are averaged on the number of images. $Image$ $Dice$ allows to the so called per-image evaluation \cite{csurka2013good}.
Per-image evaluations are important because they tend to highlight the local behaviour. For our comparisons, we calculate both $Dice$ and $Image$ $Dice$.

A score similar to $Dice$ is the Intersection Over Union ($IoU$): 
\begin{equation}
IoU = \frac{TP}{TP+FP+FN}
\end{equation}
where the difference is in the weight of $TP$.

The $F1$ Score (calculated for whole objects and not for single voxels) is defined as:
\begin{equation}
F1 = 2*\frac{OSENS * OPPV}{OSENS + OPPV}
\end{equation}
where $OSENS$ and $OPPV$ are defined above.


The Boundary F1 score ($BF$) is a per-image version of $F1$ score.

Pearson Correlation Coefficient ($PCC$), between two data sets $A$ and $B$, is defined as:
\begin{equation}
PCC(A,B) = \frac{cov(A,B)}{\sigma_{A} * \sigma_{B}}
\end{equation}
where $cov(A,B)$ is the covariance of $A$ and $B$ and $\sigma_{A}$ and $\sigma_{B}$ are the standard deviation of $A$ and $B$, respectively. $PCC$ ranges in the interval $[-1,1]$ and a negative value of $PCC$ indicates a similarity of the object $A$ with the negative version of the object $B$.

\subsection{Metrics (convergent to 0)}
The following metrics are those used in the present manuscript in which the ideal value is 0. 

Extra Fraction ($EF$), is defined as:
\begin{equation}
	EF = \frac{FP}{TP+FN}
\end{equation}
Detection error rate ($DER$) is defined as:
\begin{equation}
DER = \frac{DE}{MTA}
\end{equation}
where the detection error $DE$ is calculated as the sum of the voxels of connected regions wrongly marked as positive by the rater and the mean total area $MTA$ is defined as the average between the number of positive voxels from the rater and the ground-truth. 
$DER$ measures the disagreement in detecting the same regions between a rater under and the ground-truth.

Outline Error Rate ($OER$) is defined as:  
\begin{equation}
OER = \frac{OE}{MTA}
\end{equation}
where $OE$ is the outline error calculated as the difference between the number of voxels of 
the union and that of the intersection between the positive connected regions of both the rater and the ground-truth. $OER$ measures the disagreement in outlining the same objects between a rater and the ground truth.

False Detection Ratio ($FDE$) is defined as: 
\begin{equation}
FDE = \frac{FP}{P}
\end{equation}

Relative Area Error ($RAE$) is defined as:
\begin{equation}
RAE = \frac{TP+FP-P}{P}
\end{equation}


Hausdorff Distance ($HD$) between two objects $A$ and $B$ is defined as:

\begin{equation}
	HD(A, B) = max(h(A, B), h(B, A))
\end{equation}
where $h(A, B)$ is:
\begin{equation}
	h(A,B) = max_{a \in A} min_{b \in B} \parallel a-b \parallel
\end{equation}
$HD$ measures how far two subsets are from each other. In other words, two sets are close with respect to $HD$ if every point of one set is close to a certain point of the other set.

Euclidean Distance ($ED$) between two objects $A$ and $B$ is defined as:
\begin{equation}
ED(A, B) = max(d(A, B), d(B,A))
\end{equation}
where $d(A, B)$ is defined as:
\begin{equation}
d(A, B) = \frac{1}{N} \sum_{a \in A}^{} min_{b \in B} \parallel a - b \parallel
\end{equation}

Surface Distance ($SD$) is defined as:
\begin{equation}
SD = \frac{\sum_{i \in A_{S}} d(x_{i}, G_{S}) + \sum_{j \in G_{S}} d(x_{j}, A_{S}) } {N_{A} + N_{G}}  
\end{equation}

where $A_{S}$ and $G_{S}$ are two segmentations (one is the rater segmentation and the other is the ground truth), $d$ denotes the minimal $ED$ between voxels on both surfaces, while $N_{A}$ and $N_{G}$ denote the number of points of each surface.

\section{Results and Discussion}

The proposed framework has been trained, validated and tested on the ternary ground-truth defined above. As far as the ground-truth maintains unaltered Lesion of the original binary ground-truth, we guarantee a direct comparison with human raters and, at the same time, with already existing automated methods.
Regarding the segmented Uncertainty, a comparison is possible just with respect to the ternary ground-truth, since for the human raters Uncertainty is unavailable. 

The evaluation of the proposed framework and of the human radiologists, with respect both to the ground-truth and to each-other, is performed by applying the cross-validation approach defined in Subsection \ref{subsecMSSEG}. Average and standard deviation values are calculated for the evaluation parameters defined in Section \ref{sec:perf} and divided in two groups: those whose ideal value is 1 and those whose ideal value is 0. 

\begin{figure}[!t]
\centering
\includegraphics[scale=0.38]{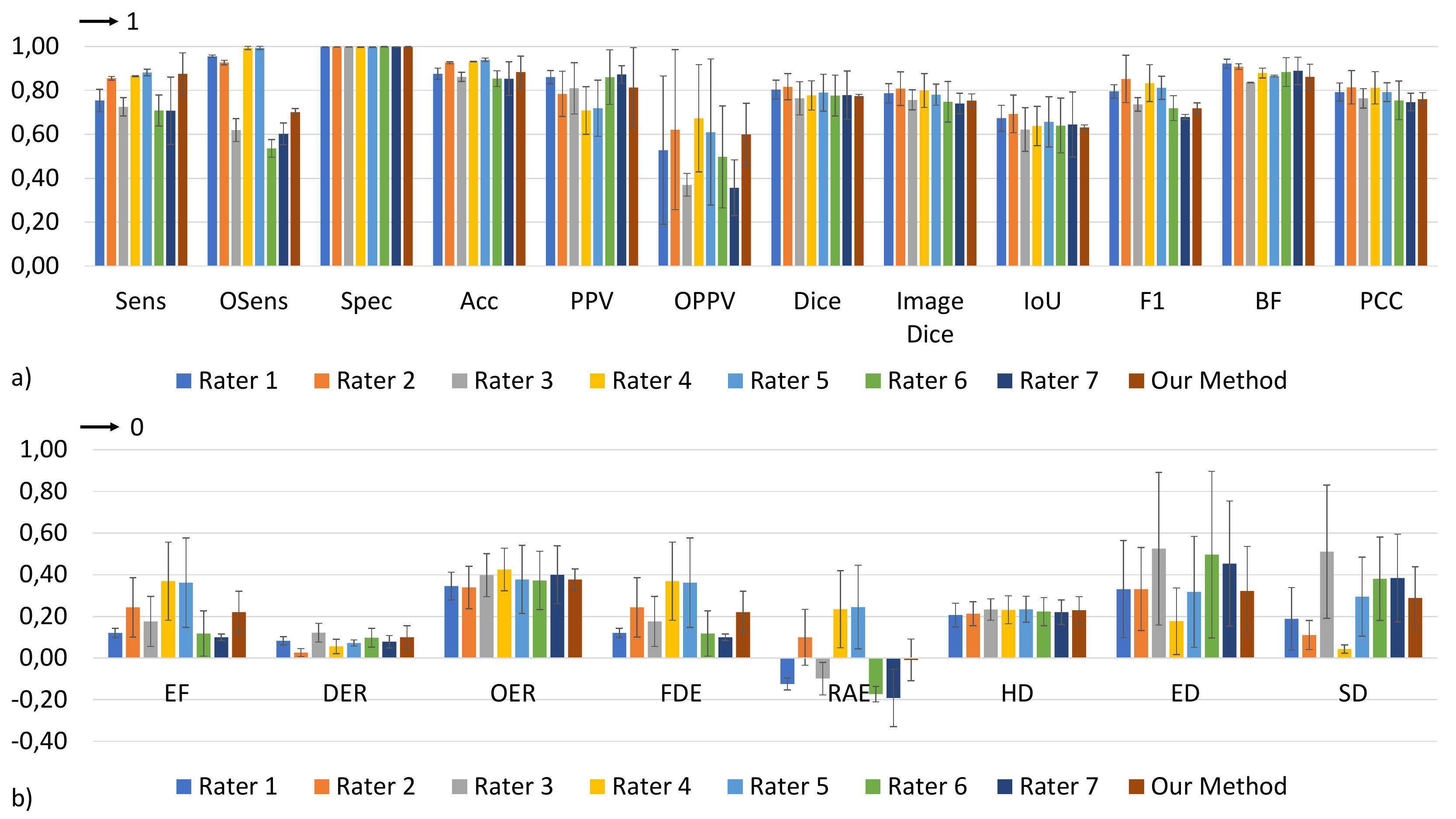}
\caption{\label{figstic}Comparison between the raters and the ground truth, performed on the Lesion class. The reported metrics are separated in those whose ideal value is 1 (a) and those whose ideal value is 0 (b). Average and standard deviation are reported. Euclidean, Hausdorff and Surface distances are shown in cm units.}
\end{figure}

\begin{figure}[!t]
\centering
\includegraphics[scale=0.38]{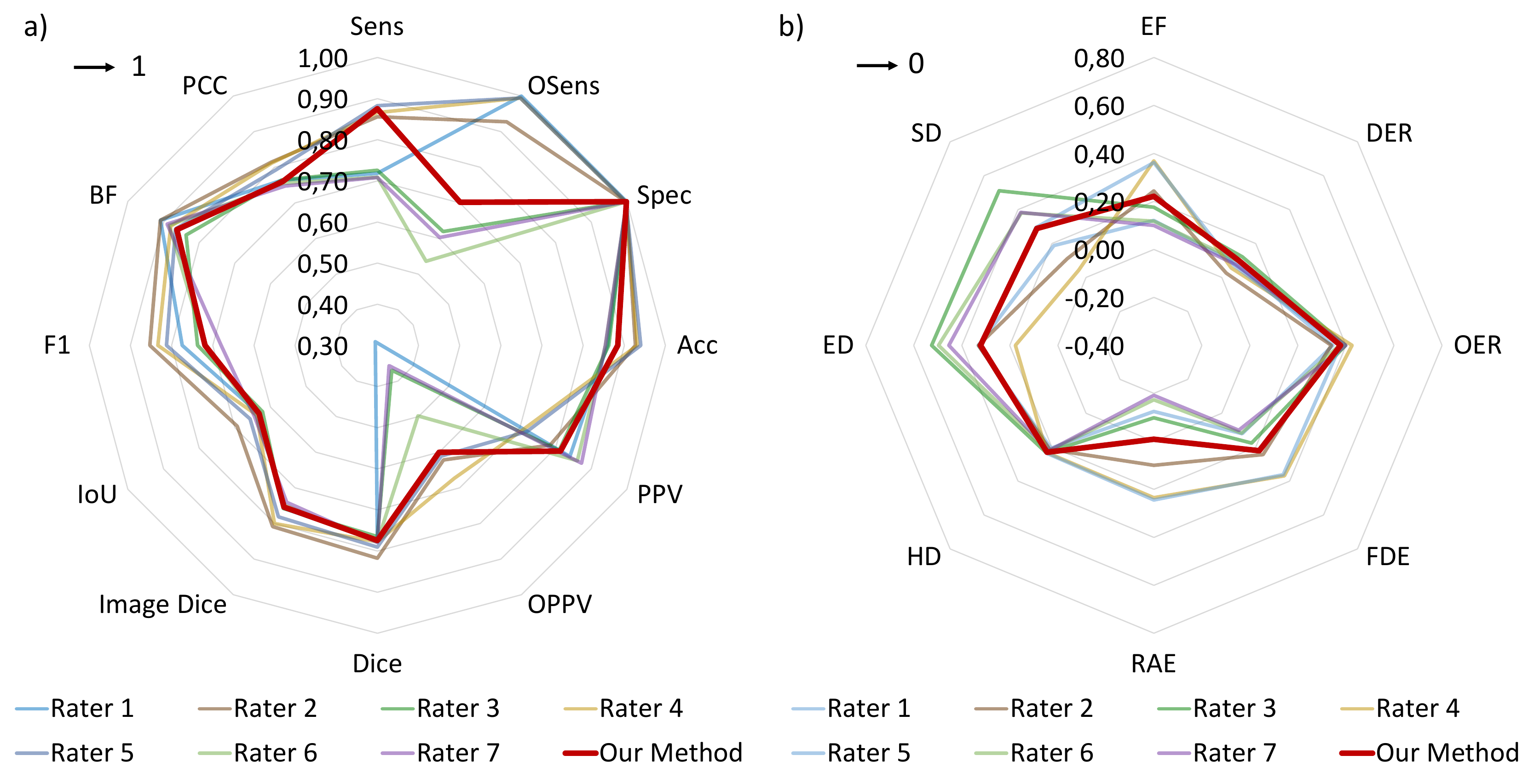}
\caption{\label{figradar}Comparison between raters in the same conditions of Fig. \ref{figstic}. For graphic purposes, only the average values are reported and the line of the proposed framework (red) is highlighted with respect to the others.}
\end{figure}

The first results, reported in Fig \ref{figstic}, are those between the raters and the ground-truth performed on the Lesion class. This is also an indirect comparison, through the ground-truth, between the proposed framework and the human raters. 
For a better overview, the mean values are also shown in Fig. \ref{figradar} by using a radar visualization: they confirm that the behaviour of the proposed method is inside the inter-rater variability. 

As it can be observed, our framework is never the best or the worst, for at least one of the metrics, as instead occurs for human raters. This can be explained, at least partially, by the fact that it has been trained with the consensus that, for its nature, tends to average pros and cons of the raters from which it has been derived. 

A Wilcoxon signed-rank test of the vectors of metric values confirms, with a significance level of 0.01, that there is no significant difference between the behaviour of our method and that of the 7 human raters, with respect to the ground-truth, on the Lesion class.
This means that, if data are shown without labels, it would be impossible to recognize the automated rater from humans.

\begin{figure}[!t]
\centering
\includegraphics[scale=0.55]{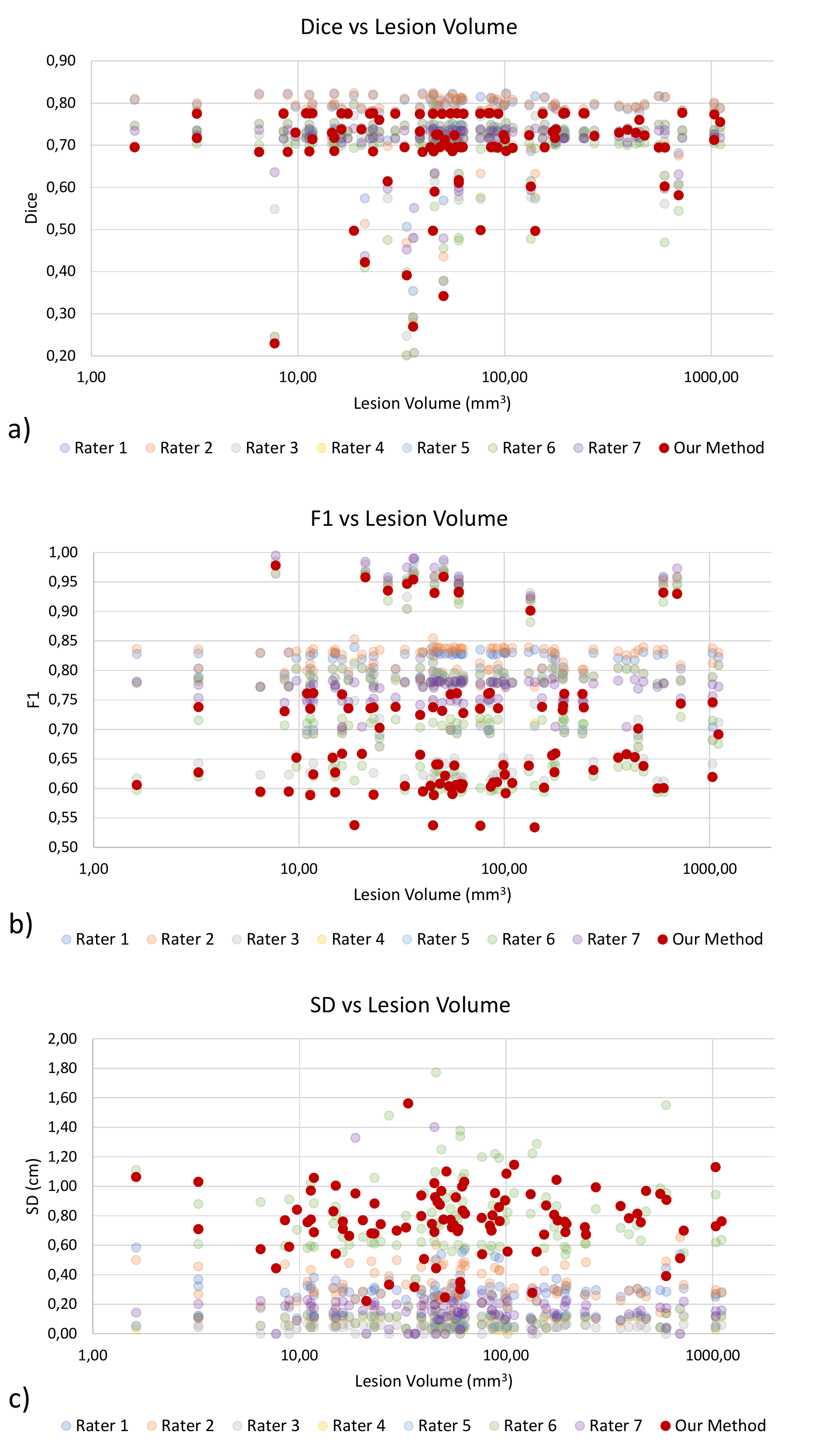}
\caption{\label{figlesvol}Dice score (a), F1-score (b) and Surface Distance (c), calculated for each lesion and shown with respect to the lesion volume for the human raters and the proposed framework. To improve readability; lesion volume is represented in logarithmic scale; framework's values (red) are highlighted.}
\end{figure}

As far as the lesion size could greatly affect the performance of the classification \cite{Commowick2018} and the previously reported results are averaged with respect to the lesion volume, we repeat the comparison by changing the lesion volume. To this aim, we consider all the lesions separately and the calculations are performed lesion by lesion, by maintaining separated also the lesions of the same volume. In this way we can: 1) visualize potential outliers; 2) represent lesion density; 3) avoid local averaging that could mask specific contributions. The results, reported just for the most commonly used evaluation parameters \cite{Commowick2018}, are shown in Fig. \ref{figlesvol}.   

It is confirmed the analogy of behavior between the proposed framework and the 7 human raters, though for some volumes our framework has exhibited results close to the borders of oscillation of the 7 human raters. Indeed, a greater dispersion can be observed for F1-score and a relatively high average value is observable for $SD$, though in line with that of some human raters. 

\begin{figure}[!t]
\centering
\includegraphics[scale=0.55]{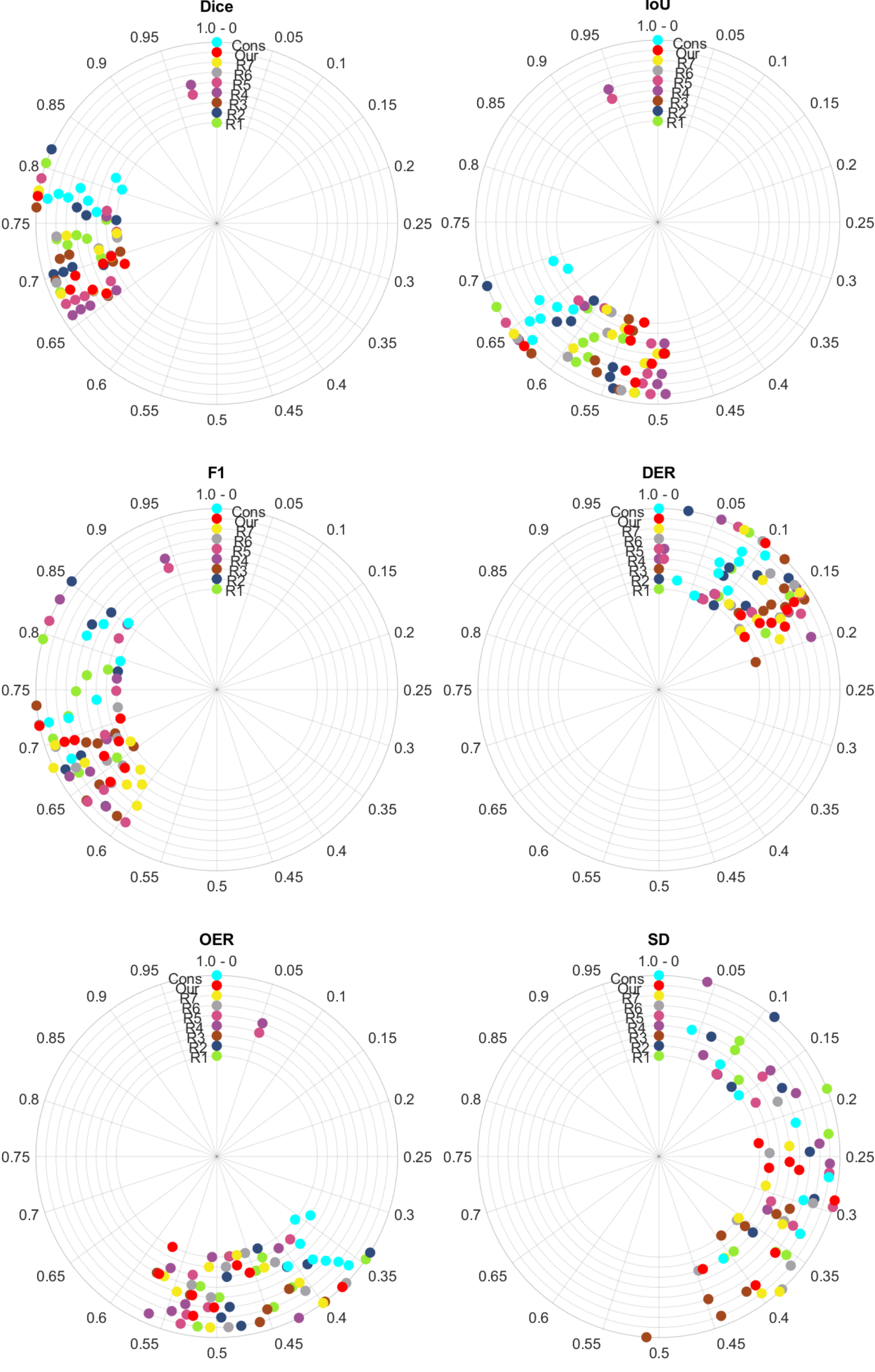}
\caption{\label{figpol}Comparison between all the raters with respect to each other, including our framework and consensus, each alternately considered as the ground-truth. For readability purposes, only some metrics are reported. 
In this representation, placing together metrics converging to 1 and to 0, the angular position indicates the metric value: clock wise versus for metrics converging to 1, anti-clock wise versus for metrics converging to 0. Radial information is only used to separate the current ground-truth raters. Ground-truth raters have colored bullets placed on the vertical line. }
\end{figure}

The above good results are not sufficient alone to state 
that our framework behaves like human raters because the comparison is through the ground-truth. In other words, our framework could be at the same 'distance' as the human raters are from the ground-truth, but from opposite sides. For this reason, a direct comparison is necessary to finally confirm the similarity between the proposed framework and the human raters. To this aim, we perform the experiment of comparing all the raters to each other by considering ground-truth all of them, including our framework, in rotation. Results, for the most frequently used metrics, are reported in Fig. \ref{figpol}.

These results confirm that the proposed framework behaviour does not differ from that of the other human raters and that it is not polarized toward a specific rater or toward the ground-truth. Moreover, as other Authors have highlighted \cite{gros2021softseg}, results show similarities among some human raters (R4 with R5 and R6 with R7). Fortunately, results also confirm that the ground-truth is not biased by the similarity between some raters, and that it maintains a "human" behaviour, being it very close to the raters R1 and R2. This is a fundamental aspect because it means that all the people who are attempting to train automated systems with respect to the ground-truth, including ourselves, are not following a "chimera" to which, paradoxically, the closer we get, the more we move away from the proper objective.
%

\begin{figure}[!t]
\centering
\includegraphics[scale=0.46]{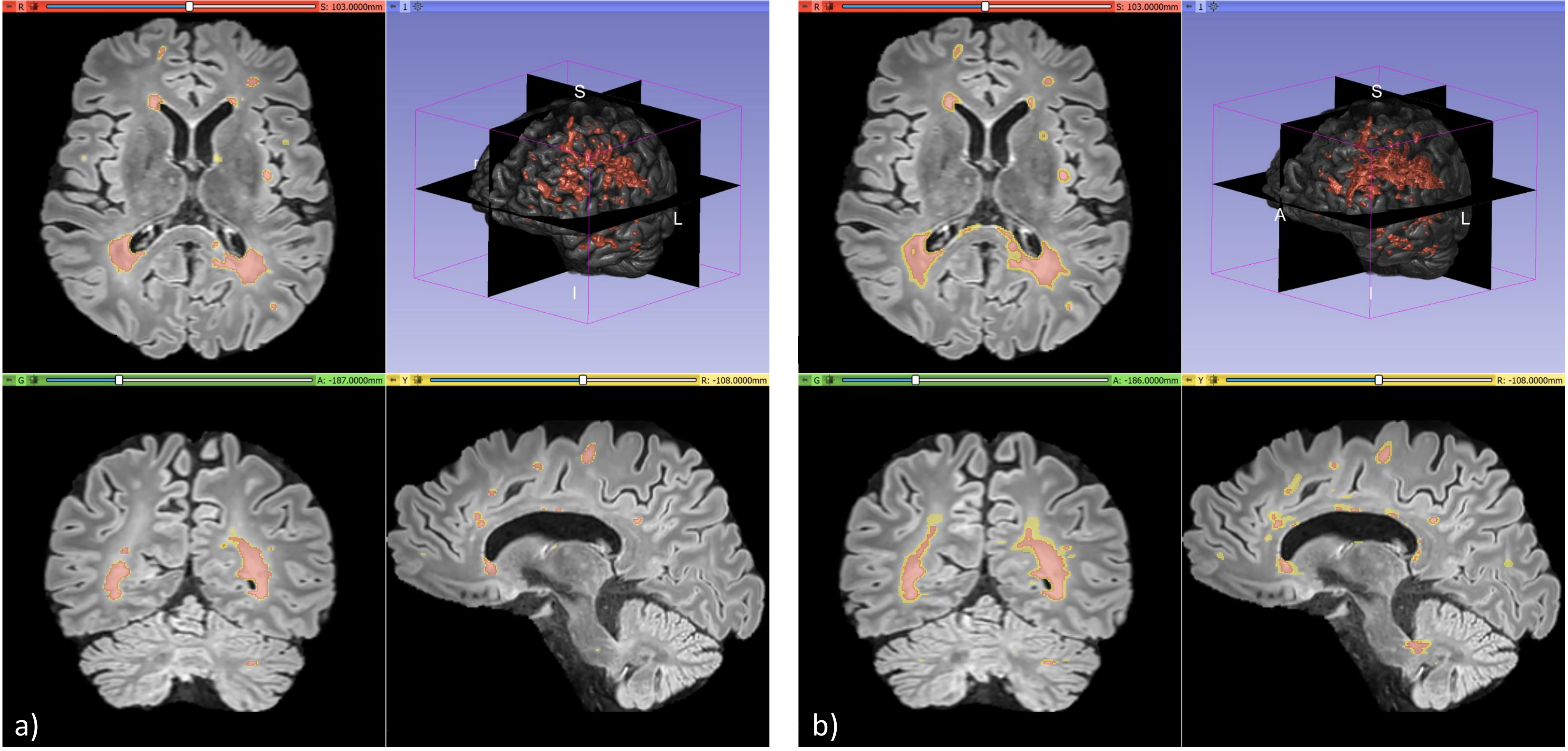}
\caption{\label{figvolume}Comparison between the ternary ground truth (left) and the proposed automated framework (right). Lesion is red and Uncertainty is yellow. For readability purposes, the upper right panel of each side shows just the healthy brain and Lesion in 3D.}
\end{figure}

A visual overview of the behaviour of the proposed framework in the whole process of identification/segmentation, both for Lesion and for Uncertainty, with respect to the ternary ground truth, is shown in Fig. \ref{figvolume}. The ternary ground-truth is reported on the left side, the corresponding segmentation obtained with the proposed framework is presented on the right side, for the same subject and slices. Both for Lesion and Uncertainty, the proposed framework selects more than necessary ($FP$ are evident). Interestingly, $FN$ are almost absent from the segmented volume. The other interesting property shown by the proposed framework is the good spatial continuity of the lesion structures in the 3D model of Lesion (upper right panel, where Lesion is red colored while Uncertainty is yellow colored).

To complete our discussion, we compare the proposed framework with recently proposed automated strategies. To this aim, Table \ref{tabcpw} contains this indirect comparison on the metrics calculated for at least one of the other methods. The necessary condition for a method to be considered in Table \ref{tabcpw} is to have been trained, validated and tested on the MSSEG data set. In this way, we can ensure that the comparison is homogeneous and performed on the same conditions of that obtained with respect to the 7 human raters. 

\begin{table}[]
\centering
\caption{\label{tabcpw}Comparison between our framework and the state of the art methods. Just averages are reported and the symbol '-' is for unavailable data. For the state of the art strategies, the performance values are those reported in the original manuscripts, while for Team Fusion (TF) the metrics are those reported in \cite{Commowick2018}.}

\tabcolsep=0.05cm
\scalebox{0.94}{
\renewcommand{\arraystretch}{2}
\begin{tabular}{ccccccccccc}
\hline
\textbf{Method} &
  \textbf{MRI mod.} &
  \textbf{Sens} &
  \textbf{OSens} &
  \textbf{TPR} &
  \textbf{Acc} &
  \textbf{PPV} &
  \textbf{OPPV} &
  \textbf{Dice} &
  \textbf{F1} &
  \textbf{SD} \\ \hline
\renewcommand{\arraystretch}{1}
\begin{tabular}[c]{@{}c@{}}TS \cite{Commowick2018}\end{tabular} &
\renewcommand{\arraystretch}{1}
\begin{tabular}[c]{@{}c@{}}FLAIR, PD, T2\\ T1, G-E T1\end{tabular} &
  0.71 & 0,60 & 0.99 & - & 0.65 & 0.53 & 0.64 & 0.50 & 0.91 \\
\cite{ghosal2019light}          &
\renewcommand{\arraystretch}{1}
\begin{tabular}[c]{@{}c@{}}FLAIR, PD, T2\\ T1, G-E T1\end{tabular} & 0.65 & -    & 0.86 & 0.97 & -    & -    & 0.76 & -    & -    \\
\cite{Valverde2019}             & FLAIR, T1                                                          & 0.55 & -    & -    & -     & -    & 0.79 & 0.63 & -    & -    \\
\cite{alijamaat2021multiple}    &
\renewcommand{\arraystretch}{1}
\begin{tabular}[c]{@{}c@{}}FLAIR, PD\\ T1, G-E T1\end{tabular}     & 0.76 & -    & -    & -     & -    & -    & 0.82 & -    & -    \\
\cite{mckinley2021simultaneous} & FLAIR, T1, T2                                                      & -    & -    & -    & -     & -    & -    & 0.76 & 0.59 & -    \\
Our                   & FLAIR                                                              & 0.88 & 0.77 & 0.98 & 0.88  & 0.81 & 0.81 & 0.77 & 0.72 & 0.27 \\ \hline
\end{tabular}
}
\end{table}

Though a global ranking is difficult, data reported in Table \ref{tabcpw} are clear: the proposed framework is the most stable with respect to different metrics and it generally outperforms the other methods, including those using multiple imaging. This could have an interesting implication: 
FLAIR would contain sufficient information, not only necessary, to identify and segment all MS lesions occurring in the WM. 
Potentially positive consequences are: a) due to the huge variability of MRI and of its modalities, the usage of a single modality could increase the performance, above using multiple modalities, because it could greatly contribute to stabilize automatic identification/classification; b)
the acquisition time and stress for the patient can be reduced.

\begin{figure}[!t]
\centering
\includegraphics[scale=0.47]{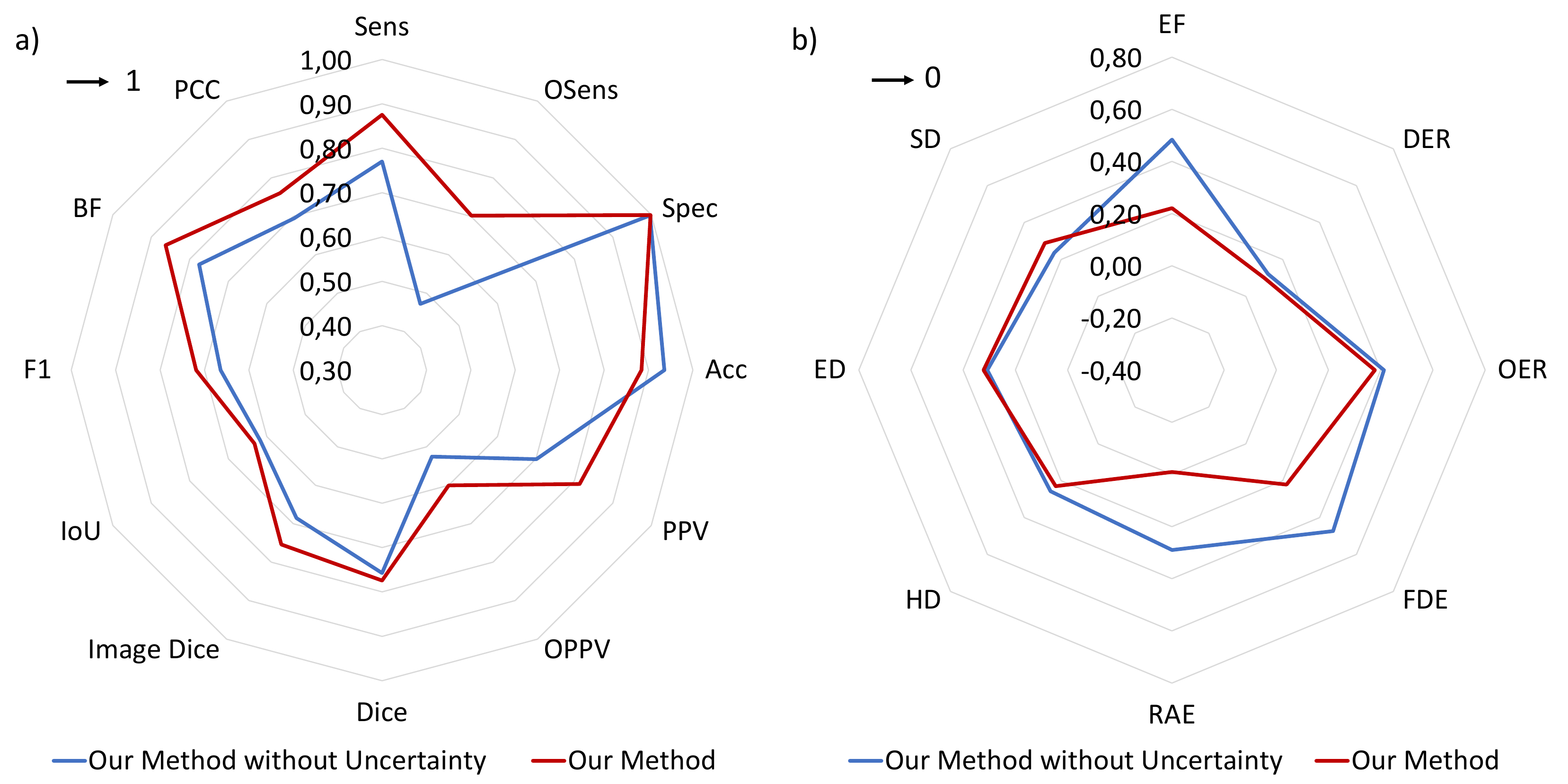}
\caption{\label{figradar_uncertainty}The proposed method when trained without and with Uncertainty, compared both on metrics whose ideal value is 1 (a) and for metrics whose ideal value is 0 (b).}
\end{figure}

An important aspect that has determined the outstanding performance of the proposed framework is the use of the ternary ground-truth. Indeed, Fig. \ref{figradar_uncertainty} shows the average performance results when the proposed framework is trained without the Uncertainty (on the binary consensus) as compared to those obtained when trained on the ternary consensus. The ensemble method trained without the Uncertainty outperforms similar automated strategies (team fusion in \cite{Commowick2018}), though it is still far from humans: the step which places the proposed framework among humans is the inclusion in the pipeline of the class Uncertainty. 

This is in line with what reported in 
\cite{gros2021softseg}: the framework learns better what is surely Lesion, what is surely Background and uses Uncertainty for doubts, as a buffer class. Indeed, the polarized and ambiguous classification of doubtful voxels, sometimes as Lesion and others as Background, disorients any automated strategy and deviates it from the human reasoning. 

\section{Conclusion and future work}
An automated framework for the identification/segmentation of MS lesions from FLAIR MRI images has been presented. We have demonstrated that traditional CNN architectures, if placed in a context emulating the procedures of human specialists, could effectively behave like a human expert. The strength points of the proposed framework are the following: 1) to train the system to recognize both the lesions as they are and with respect to the environment they are immersed in, thus allowing to incorporate also a sort of meta-information regarding the environment where MS lesions mostly occur; 2) to resemble radiologists in consulting axial slices to discover potential lesions and to check radial and sagittal slices for confirmation, as well as to maintain 3D continuity to their findings; 3) to use an ensemble classification that usually performs better than its components; 4) to use an artificially generated Uncertainty class to improve the performance of an automated strategy and to make it more similar to the human reasoning; 5) to operate just on FLAIR images.

Results have shown that the proposed framework resembles human raters both in behaviour and in performance, when compared with the MSSEG consensus on Lesion. Indeed, Wilcoxon statistical test has assessed the framework ability to exhibit a behaviour that is equivalent to, or indistinguishable from, that of a human rater. Furthermore, the proposed framework outperforms the state of the art strategies which have been trained, validated and tested on the MSSEG data set.
Finally, the usage of the uncertainty during training has greatly improved the performance of the framework.

In a recent report \cite{Jason2017}, the JASON Advisory Group has identified several key recommendations for advancing computation technology into routine clinical practice. One of them is that new technologies should address a significant clinical need, be practical in use and reduce medical system costs. The demonstration that a better performance is possible by including some concepts (Uncertainty and Ensemble) to enrich traditional CNN architectures goes in the direction of the above recommendations. 

Future directions of exploration will be: 
1) an accurate definition of "Uncertainty" and a deep study of its role in the intrinsically uncertain decision-making process;
2) the use of specific pre-processing strategy for the FLAIR images 
to further improve the robustness of the method with respect to MRI and MRI scanners variability;
3) how multiple imaging could affect the classification performance; 
4) how the use of a different loss function
could better deal the unbalancing problem 
5) to study a "soft" consensus based on a single class (lesion) with different probability values
to reduce problem complexity and to explore a "soft" loss function;
6) 
to test the proposed framework to identify/segment also cortical lesions and for longitudinal studies (temporal evolution of the disease);
7) to explore specific pre-processing strategies to make all the used modalities robust with respect to MRI and MRI scanners; 
8) to study the role of additional meta-information information
in improving lesion identification.

\section*{Acknowledgments}
The Authors wish to thank Dr Maria Silvia Marottoli for her contribution to the improvement of the manuscript.
	
	\bibliography{main}
	
\end{document}